\documentclass[journal]{IEEEtran}

\usepackage[cmex10]{amsmath}
\usepackage{cite}

\usepackage{latexsym}
\usepackage{graphics}
\usepackage{amssymb}
\usepackage{amsthm}
 \usepackage{relsize}

\usepackage{cite}
\usepackage{array}
\usepackage{mdwmath}
\usepackage{mdwtab}
\usepackage[tight,footnotesize]{subfigure}
\usepackage{graphicx}

\usepackage{stfloats}
\fnbelowfloat
\usepackage{url}

\def\SNR{\mbox{\scriptsize\sf SNR}}

\def\tr{\mathrm{tr}}

\def\rank{\mathrm{rank}}

\def\diag{\mathrm{diag}}
\def\projUpsilon{\boldsymbol{\Pi}\!\!\raisebox{-1pt}{ \scriptsize $\vect{\Upsilon}_t^{H/2} \vect{U}_M$}}

\newcommand{\vect}[1]{\mathbf{#1}}

\theoremstyle{remark}

\newtheorem{theorem}{Theorem}
\newtheorem{corollary}{Corollary}
\newtheorem{lemma}{Lemma}
\newtheorem{definition}{Definition}

\begin{document}

\title{Capacity Limits and Multiplexing Gains of\\ MIMO Channels with Transceiver Impairments}

\author{Emil Bj\"ornson, Per Zetterberg, Mats Bengtsson, and Bj\"orn~Ottersten
\thanks{\copyright \, 2013 IEEE. Personal use of this material is permitted. Permission from IEEE must be obtained for all other uses, in any current or future media, including reprinting/republishing this material for advertising or promotional purposes, creating new collective works, for resale or redistribution to servers or lists, or reuse of any copyrighted component of this work in other works.}
\thanks{Manuscript received September 5, 2012. The associate editor coordinating
the review of this letter and approving it for publication was D.-A.~Toumpakaris. Supplementary downloadable material is available at https://github.com/emilbjornson/capacity-limits-transceiver-impairments, provided by the authors. The material includes Matlab code that reproduces all simulation results.}
\thanks{This work was supported by International Postdoc Grant 2012-228 from
The Swedish Research Council and by the HIATUS project (FET 265578).}
\thanks{The authors are with the Signal Processing Lab, ACCESS Linnaeus Center,
KTH Royal Institute of Technology, SE-100 44 Stockholm, Sweden. E.~Bj\"ornson is also the Alcatel-Lucent Chair on Flexible Radio, SUPELEC 91192
Gif-sur-Yvette, France (e-mail: emil.bjornson@ee.kth.se). B.~Ottersten is also
with the Interdisciplinary Centre for Security, Reliability and Trust (SnT),
University of Luxembourg, L-1359 Luxembourg-Kirchberg, Luxembourg. Digital Object Identifier 10.1109/LCOMM.2012.112012.122003}
}

\markboth{IEEE COMMUNICATIONS LETTERS, VOL.~17, NO.~1, JANUARY 2013}%
{Bj\"ornson \MakeLowercase{\textit{et al.}}: IEEE COMMUNICATIONS LETTERS}

\maketitle

\begin{abstract}
The capacity of \emph{ideal} MIMO channels has a high-SNR slope that equals the minimum of the number of transmit and receive antennas. This letter analyzes if this result holds when there are distortions from physical transceiver impairments. We prove analytically that such \emph{physical} MIMO channels have a finite upper capacity limit, for any channel distribution and SNR. The high-SNR slope thus collapses to zero. This appears discouraging, but we prove the encouraging result that the \emph{relative} capacity gain of employing MIMO is at least as large as with ideal transceivers.
\end{abstract}


\IEEEpeerreviewmaketitle

\section{Introduction}

In the past decade, a vast number of papers have studied multiple-input multiple-output (MIMO) communications motivated by the impressive capacity scaling in the high-SNR regime. The seminal article \cite{Telatar1999a} by E.~Telatar shows that the MIMO capacity with channel knowledge at the receiver behaves as $M \log_2(\SNR) + \mathcal{O}(1)$, where $\SNR$ is the signal-to-noise ratio (SNR). The slope $M$ satisfies $M=\min(N_t,N_r)$, where $N_t$ and $N_r$ are the  number of transmit and receive antennas, respectively. $M$ is the asymptotic gain over single-antenna channels and is called \emph{degrees of freedom} or \emph{multiplexing gain}.

Some skepticism concerning the applicability of these results in cellular networks has recently appeared; modest gains of network MIMO over conventional schemes have been observed and the throughput might even decrease due to the extra overhead \cite{Barbieri2012a,Lozano2013a}. One explanation is the finite channel coherence time that limits the resources for channel acquisition \cite{Jose2011b} and coordination between nodes \cite{Lozano2013a}, thus creating a finite fundamental ceiling for the network spectral efficiency---irrespectively of the power and the number of antennas.

While these results concern large network MIMO systems, there is another non-ideality that also affects performance and manifests itself for MIMO systems of any size: \emph{transceiver impairments} \cite{Koch2009a,Schenk2008a,Studer2010a,Moghadam2012a,Holma2011a,Bjornson2012b}. Physical radio-frequency (RF) transceivers suffer from amplifier non-linearities, IQ-imbalance, phase noise, quantization noise, carrier-frequency and sampling-rate jitter/offsets, etc. These impairments are conventionally overlooked in information theoretic studies, but this letter shows that they have a non-negligible and fundamental impact on the spectral efficiency in modern deployments with high SNR.

This letter analyzes the generalized MIMO channel with transceiver impairments from \cite{Studer2010a}. We show that the capacity has a finite high-SNR limit for any channel distribution. The multiplexing gain is thus zero, which is fundamentally different from the ideal case in \cite{Telatar1999a} (detailed above). Similar single-antenna results are given in \cite{Koch2009a}. The practical  MIMO  gain---the \emph{relative} capacity increase over single-antenna channels---is however shown to be at least as large as with ideal transceivers.

\section{Generalized Channel Model}

Consider a flat-fading MIMO channel with $N_t$ transmit antennas and $N_r$ receive antennas. The received signal $\vect{y} \in \mathbb{C}^{N_r}$ in the classical affine baseband channel model of \cite{Telatar1999a} is
\begin{equation} \label{eq_channel_model}
\vect{y}= \sqrt{\SNR} \vect{H} \vect{x} + \vect{n},
\end{equation}
\noindent where $\SNR$ is the SNR, $\vect{x} \in \mathbb{C}^{N_t}$ is the intended signal, and $\vect{n} \sim \mathcal{CN}(\vect{0},\vect{I})$ is circular-symmetric complex Gaussian noise.
The channel matrix $\vect{H} \in \mathbb{C}^{N_r \times N_t}$ is assumed to be a random variable $\mathbb{H}$ having any multi-variate distribution $f_{\mathbb{H}}$ with normalized gain $\mathbb{E}\{ \tr(\vect{H}^H \vect{H} )\}=N_t N_r$ and full-rank realizations (i.e., $\rank(\vect{H})=\min(N_t,N_r)$) almost surely---this basically covers all physical channel distributions.

The intended signal  $\vect{x}$ in \eqref{eq_channel_model} is only affected by a multiplicative channel transformation and additive thermal noise, thus ideal transceiver hardware is implicitly assumed. Physical transceivers suffer from a variety of impairments that are not properly described by \eqref{eq_channel_model} \cite{Koch2009a,Schenk2008a,Studer2010a,Moghadam2012a,Holma2011a,Bjornson2012b}.
The influence of impairments is reduced by compensation schemes, leaving a residual distortion with a variance that scales with $\SNR$ \cite{Studer2010a}.

\begin{figure}
\begin{center}
\includegraphics[width=.9\columnwidth]{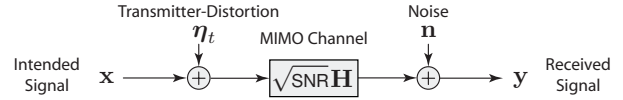}
\end{center} \vskip-4mm
\caption{Block diagram of the generalized MIMO channel considered in this letter. Unlike the classical channel model in \cite{Telatar1999a}, the transmitter distortion generated by physical transceiver implementations is included in the model.}\label{figure_block-models}
\end{figure}

A generalized MIMO channel is proposed in \cite{Schenk2008a,Studer2010a} and verified by measurements. The combined (residual) influence of impairments in the transmitter hardware is modeled by the \emph{transmitter distortion} $\boldsymbol{\eta}_{t} \in \mathbb{C}^{N_t}$ and \eqref{eq_channel_model} is generalized to
\begin{equation} \label{eq_channel_model_impairments}
\vect{y}= \sqrt{\SNR} \, \vect{H}\left( \vect{x} + \boldsymbol{\eta}_{t} \right) + \vect{n}.
\end{equation}
Note that $\boldsymbol{\eta}_{t}$ is the mismatch between the intended signal $\vect{x}$ and the signal actually radiated by the transmitter; see Fig.~\ref{figure_block-models}. It is well-modeled as uncorrelated Gaussian noise as it is the aggregate residual of many impairments, whereof some are Gaussian and some behave as Gaussian when summed up \cite{Studer2010a}.

Under the normalized power constraint\footnote{The power constraint is only defined on the intended signal, although distortions also contribute a small amount of power. However, this extra power is fully characterized by the SNR and we therefore assume that $\SNR$ is selected to make the total power usage fulfill all external system constraints.} $\tr(\vect{Q}) = 1$ with $\vect{Q}=\mathbb{E}\{ \vect{x}\vect{x}^H \}$ (similar to \cite{Telatar1999a}), the transmitter distortion is
\begin{equation*}
\boldsymbol{\eta}_{t} \sim \mathcal{CN} \big( \vect{0}, \vect{\Upsilon}_t(\vect{Q}) \big) \!\!\! \quad \textrm{with} \quad\!\! \vect{\Upsilon}_t \!=\! \diag(\upsilon_1(q_1),\ldots,\upsilon_{N_t}(q_{N_t})).
\end{equation*}
\noindent The distortion depends on the intended signal $\vect{x}$ in the sense that the variance $\upsilon_n(q_n)$ is an increasing function of the signal power $q_n$ at the $n$th transmit antenna (i.e., the $n$th diagonal element of $\vect{Q}$). We neglect any antenna cross-correlation in $\vect{\Upsilon}_t$.\footnote{Correlation between antennas is predicted in \cite{Moghadam2012a}, but it is typically small.} In multi-carrier (e.g., OFDM) scenarios, \eqref{eq_channel_model_impairments} can describe each individual subcarrier. However, there is some distortion leakage between subcarriers that makes $q_n$ less influential on $\upsilon_n(q_n)$. For simplicity, we model the leakage as proportional to the average signal power per antenna (i.e., the direct impact of what is done on individual antennas/subcarriers averages out when having many subcarriers). To capture a range of cases we propose
\begin{equation} \label{eq_distortion_model}
\upsilon_n(q_n) = \kappa^2 \Big( \, (1\!-\!\alpha) q_n + \alpha \frac{\sum_{i=1}^{N_t} q_i}{N_t} \Big),
\end{equation}
where the parameter $\alpha \in [0,1]$ enables transition from one ($\alpha\!=\!0$) to many ($\alpha\!=\!1$) subcarriers. The parameter $\kappa>0$ is the \emph{level of impairments}.\footnote{The error vector magnitude, $\mathrm{EVM}=\frac{\mathbb{E}\{\|\boldsymbol{\eta}_{t}\|^2\} }{\mathbb{E}\{\|\vect{x}\|^2\}}$, is a common measure for quantifying RF transceiver impairments. Observe that the EVM equals $\kappa^2$ for the considered $\upsilon_n(q_n)$ in \eqref{eq_distortion_model}. EVM requirements in the range $\kappa \in [0.08,0.175]$ occur in Long Term Evolution (LTE) \cite[Section 14.3.4]{Holma2011a}.} This model is a good characterization of phase noise and IQ-imbalance, while the impact of amplifier non-linearities grows non-linearly in $\SNR$  \cite{Schenk2008a}. We assume to operate in the dynamic range where the impact is almost linear.

\section{Analysis of Channel Capacity}

The transmitter knows the channel distribution $f_{\mathbb{H}}$, while the receiver knows the realization $\vect{H}$. The capacity of \eqref{eq_channel_model_impairments} is
\begin{equation} \label{eq_capacity_initial}
C_{N_t,N_r}(\SNR) = \sup_{f_{\mathbb{X}}: \, \tr(\mathbb{E}\{ \vect{x}\vect{x}^H \}) = \tr(\vect{Q}) = 1} \mathcal{I}(\vect{x}; \vect{y} | \mathbb{H})
\end{equation}
where $f_{\mathbb{X}}$ is the PDF of $\vect{x}$ and $\mathcal{I}(\cdot ; \cdot |\cdot)$ is conditional mutual information. Note that
$\mathcal{I}(\vect{x}; \vect{y} | \mathbb{H})  = \mathbb{E}_{\vect{H}} \{\mathcal{I}(\vect{x}; \vect{y} | \mathbb{H} = \vect{H}) \}$.

\begin{lemma} \label{lemma_capacity_expression}
The capacity $C_{N_t,N_r}(\SNR)$ can be expressed as
\begin{equation*}
\sup_{\vect{Q} : \, \tr(\vect{Q}) = 1} \!\! \mathbb{E}_{\vect{H}} \Big\{ \! \log_2 \det \! \big( \vect{I} + \SNR \vect{H} \vect{Q} \vect{H}^H ( \SNR \vect{H} \vect{\Upsilon}_t \vect{H}^H \!+ \vect{I})^{\!-1} \big) \! \Big\}
\end{equation*}
and is achieved by $\vect{x} \sim \mathcal{CN}(\vect{0},\vect{Q})$ for some feasible $\vect{Q} \succeq \vect{0}$.
\end{lemma}
\begin{IEEEproof}
For any realization $\mathbb{H} = \vect{H}$ and fixed $\SNR$, \eqref{eq_channel_model_impairments} is a classical MIMO channel but with noise covariance $(\SNR \vect{H} \vect{\Upsilon}_t \vect{H}^H \!+\! \vect{I})$. The given expression and the sufficiency of using a Gaussian distribution on $\vect{x}$ follow from \cite{Telatar1999a}.
\end{IEEEproof}

Although the capacity expression in Lemma \ref{lemma_capacity_expression} looks similar to that of the classical MIMO channel in \eqref{eq_channel_model} and \cite{Telatar1999a}, it behaves very differently---particularly in the high-SNR regime.

\begin{theorem} \label{theorem_upper_bound}
The asymptotic capacity limit $C_{N_t,N_r}(\infty) = \lim_{\SNR \rightarrow \infty} C_{N_t,N_r}(\SNR)$ is finite and bounded as
\begin{equation} \label{eq_capacity_bounds}
M \log_2 \! \left( \! 1+\frac{1}{\kappa^2} \!\right) \leq C_{N_t,N_r}(\infty) \leq M \log_2 \! \left( \! 1+\frac{N_t}{M \kappa^2} \! \right)
\end{equation}
where $M \!=\! \min(N_t,N_r)$. The lower bound is asymptotically achieved by $\vect{Q} = \frac{1}{N_t} \vect{I}$. The two bounds coincide if $N_t \leq N_r$.
\end{theorem}
\begin{IEEEproof}
The proof is given in the appendix.
\end{IEEEproof}

This theorem shows that physical MIMO systems have a finite capacity limit in the high-SNR regime---this is fundamentally different from the unbounded asymptotic capacity for ideal transceivers \cite{Telatar1999a}. Furthermore, the bounds in \eqref{eq_capacity_bounds} hold for any channel distribution and are only characterized by the number of antennas and the level of impairments $\kappa$.

The bounds in \eqref{eq_capacity_bounds} coincide for $N_t \leq N_r$, while only the upper bound grows with the number of transmit antennas when $N_t > N_r$. Informally speaking, the lower and upper bounds are tight when the high-SNR capacity-achieving $\vect{Q}$ is isotropic in a subspace of size $N_t$ and size $\min(N_r,N_t)$, respectively. The following corollaries exemplify these extremes.

\begin{corollary} \label{corollary_rotationally_invariant}
Suppose the channel distribution is right-rotationally invariant (e.g., $\mathbb{H} \sim \mathbb{H} \vect{U}$ for any unitary matrix $\vect{U}$).
The capacity is achieved by $\vect{Q} = \frac{1}{N_t} \vect{I}$ for any $\SNR$ and $\alpha$.
The lower bound in \eqref{eq_capacity_bounds} is asymptotically tight for any $N_t$.
\end{corollary}
\begin{IEEEproof}
The right-rotational invariance implies that the $N_t$ dimensions of $\vect{H}^H \vect{H}$ are isotropically distributed, thus the concavity of $\mathbb{E}\{ \log \det(\cdot) \}$ makes an isotropic covariance matrix optimal. The lower bound in \eqref{eq_capacity_bounds} is asymptotically tight as it is constructed using this isotropic covariance matrix.
\end{IEEEproof}

This corollary covers Rayleigh fading channels that are uncorrelated at the transmit side, but also other channel distributions with isotropic spatial directivity at the transmitter.

The special case of a deterministic channel matrix enables stronger adaptivity of $\vect{Q}$ and achieves the upper bound in \eqref{eq_capacity_bounds}.

\begin{corollary} \label{corollary_capacity_deterministic}
Suppose $\alpha=1$ and the channel $\vect{H}$ is deterministic and full rank. Let $\vect{H}^H \vect{H} = \vect{U}_M \boldsymbol{\Lambda}_M \vect{U}_M^H$ denote a compact eigendecomposition, where $\boldsymbol{\Lambda}_M = \diag(\lambda_1,\ldots,\lambda_M)$ contains the non-zero eigenvalues and the semi-unitary $\vect{U}_M \in \mathbb{C}^{N_t \times M}$ contains the corresponding eigenvectors.
The capacity is
\begin{equation}
C_{N_t,N_r}(\SNR) = \sum_{i=1}^{M} \log_2\bigg( 1 + \frac{\SNR \lambda_i d_i}{\SNR \lambda_i \frac{\kappa^2}{N_t} + 1}  \bigg)
\end{equation}
for $d_i = \big[ \mu - \frac{1}{\lambda_i} \big]_+$ where $\mu$ is selected to make $\sum_{i=1}^{M} d_i = 1$. The capacity is achieved by $\vect{Q} = \vect{U}_M \diag(d_1,\ldots,d_M) \vect{U}_M^H$. The upper bound in \eqref{eq_capacity_bounds} is asymptotically tight for any $N_t$.
\end{corollary}
\begin{IEEEproof}
The capacity-achieving $\vect{Q}$ is derived as in \cite{Telatar1999a}, using the Hadamard inequality.
The capacity limit follows since $\vect{Q} = \frac{1}{M} \vect{U}_M \vect{U}_M^H$ achieves the upper bound in \eqref{eq_capacity_bounds}.
\end{IEEEproof}

Although the capacity behaves differently under impairments, the optimal waterfilling power allocation in Corollary~\ref{corollary_capacity_deterministic} is the same as for ideal transceivers (also noted in \cite{Studer2010a}). When $N_t \geq N_r$, the capacity limit $M \log_2 ( 1+\frac{N_t}{M \kappa^2} )$ is improved by increasing $N_t$, because a deterministic $\vect{H}$ enables selective transmission in the $N_r$ non-zero channel dimensions while the transmitter distortion is isotropic over all $N_t$ dimensions.

We conclude the analysis by elaborating on the fact that the lower bound in \eqref{eq_capacity_bounds} is always asymptotically achievable.

\begin{corollary}
If the channel distribution $f_{\mathbb{H}}$ is unknown at the transmitter, the worst-case mutual information $\min_{f_{\mathbb{H}}} \mathcal{I}(\vect{x}; \vect{y} | \mathbb{H})$
 is maximized by $\vect{Q} \!=\! \frac{1}{N_t} \vect{I}$ (for any $\alpha$) and approaches $M \log_2 \! \left( 1+\frac{1}{\kappa^2} \!\right)$ as $\SNR \!\rightarrow \!\infty$.
\end{corollary}

\subsection{Numerical Illustrations}
\label{subsection_numerical_illustration}

\begin{figure}
\begin{center}
\includegraphics[width=\columnwidth]{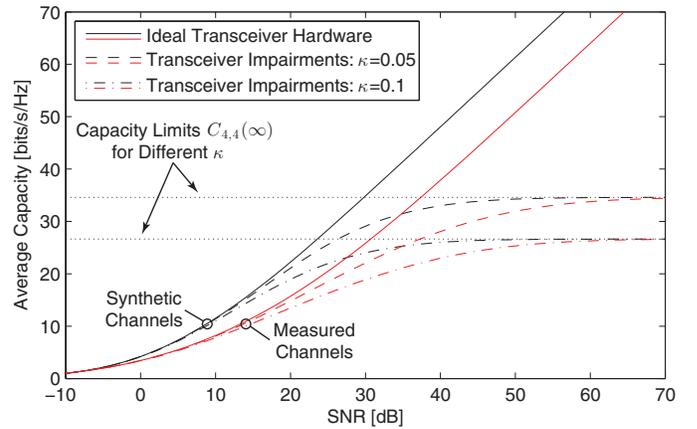}
\end{center} \vskip-4mm
\caption{Average capacity of a 4x4 MIMO channel over different deterministic channel realizations, different levels of transceiver impairments, and $\alpha=1$.}\label{figure_simulation_capacity}
\end{figure}

Consider a channel with $N_t=N_r=4$, $\alpha=1$, and varying SNR. Fig.~\ref{figure_simulation_capacity} shows the average capacity over different deterministic channels, either generated synthetically with independent $\mathcal{CN}(0,1)$-entries or taken from the measurements in \cite{Jalden2007a}. The level of impairments is varied as $\kappa \in \{ 0.05, \,0.1\}$.

Ideal and physical transceivers behave similarly at low and medium SNRs in Fig.~\ref{figure_simulation_capacity}, but fundamentally different at high SNRs. While the ideal capacity grows unboundedly, the capacity with impairments approaches the capacity limit $C_{4,4}(\infty)= 4 \log_2  ( 1+\frac{1}{\kappa^2} )$ in Theorem~\ref{theorem_upper_bound}. The difference between the uncorrelated synthetic channels and the realistically correlated measured channels vanishes asymptotically. Therefore, only the level of impairments, $\kappa$, decides the capacity limit.

\begin{figure}
\begin{center}
\includegraphics[width=\columnwidth]{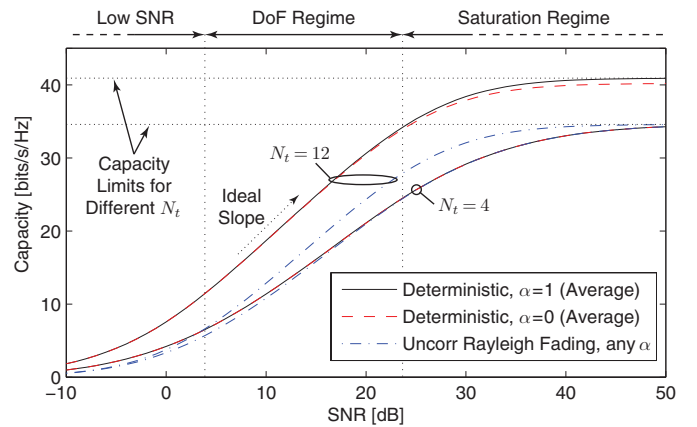}
\end{center} \vskip-4mm
\caption{Capacity of a MIMO channel with $N_r=4$ and impairments with $\kappa = 0.05$. We consider different $N_t$, channel distributions, and $\alpha$-values.}\label{figure_simulation_capacity_antennas}
\end{figure}

Next, we illustrate the case $N_t \geq N_r$ and different $\alpha$. Fig.~\ref{figure_simulation_capacity_antennas} considers $N_t \in \{4, \, 12\}$ with $N_r=4$, $\kappa = 0.05$, and two different channel distributions: deterministic (average capacity with known i.i.d. $\mathcal{CN}(0,1)$-entries) and uncorrelated Rayleigh fading. We show $\alpha \in \{0,\,1\}$ in the deterministic case, while the random case gives $\vect{Q}=\frac{1}{N_t} \vect{I}$ and same capacity for any $\alpha$.

These channels perform similarly and have the same capacity limit when $N_t=4$.
The convergence to the capacity limit becomes faster for the random distribution when $N_t$ increases, but the value of the limit is unchanged. Contrary, the capacity limits in the deterministic cases increase with $N_t$ (and with $\alpha$ since it makes the distortion more isotropic). Fig.~\ref{figure_simulation_capacity_antennas} shows that there is a medium SNR range where the capacity exhibits roughly the same $M$-slope as achieved asymptotically for ideal transceivers. Following the terminology of \cite{Lozano2013a}, this is the \emph{degrees-of-freedom (DoF) regime} while the high-SNR regime is the \emph{saturation regime}; see Fig.~\ref{figure_simulation_capacity_antennas}. This behavior appeared in \cite{Lozano2013a} for large cellular networks due to limited coherence time, but we demonstrate its existence for any physical MIMO channel (regardless of size) due to transceiver impairments.

\section{Gain of Multiplexing}

The MIMO capacity with ideal transceivers behaves as $M \log_2(\SNR) + \mathcal{O}(1)$ \cite{Telatar1999a}, thus it grows unboundedly in the high-SNR regime and scales linearly with the so-called multiplexing gain $M = \min(N_t,N_r)$. On the contrary, Theorem~\ref{theorem_upper_bound} shows that the capacity of physical MIMO channels has a finite upper bound, giving a very different multiplexing gain:
\begin{equation} \label{eq_classic_multiplexing_gain}
\mathcal{M}_{\infty}^{\textrm{classic}} = \lim_{\SNR \rightarrow \infty} \frac{C_{N_t,N_r}(\SNR)}{\log_2 ( \SNR )} = 0.
\end{equation}

In view of \eqref{eq_classic_multiplexing_gain}, one might think that the existence of a non-zero multiplexing gain is merely an artifact of ignoring the transceiver impairments that always appear in practice. However, the problem lies in the classical definition, because also physical systems can gain in capacity from employing multiple antennas and utilizing spatial multiplexing.
A practically more relevant measure is the relative capacity improvement (at a finite $\SNR$) of an $N_t \times N_r$ MIMO channel over the corresponding single-input single-output (SISO) channel.

\begin{definition} \label{definition_finiteSNR_multiplexing}
The \emph{finite-SNR multiplexing gain}, $\mathcal{M}(\SNR)$, is the ratio of MIMO to SISO capacity at a given $\SNR$. For \eqref{eq_channel_model_impairments},
\begin{equation} \label{eq_refined_multiplexing_gain}
\mathcal{M}(\SNR) = \frac{C_{N_t,N_r}(\SNR)}{C_{1,1}(\SNR)}.
\end{equation}
\end{definition}

This ratio between the MIMO and SISO capacity quantifies the exact gain of multiplexing.
The concept of a finite-SNR multiplexing gain was introduced in \cite{Narasimhan2006a} for ideal transceivers, while the refined Definition \ref{definition_finiteSNR_multiplexing} can be applied to any channel model. The asymptotic behavior of $\mathcal{M}(\SNR)$ is as follows.

\begin{theorem} \label{theorem_finite_multiplexing_gain}
Let $h$ denote the SISO channel. The finite-SNR multiplexing gain, $\mathcal{M}(\SNR)$, for \eqref{eq_channel_model_impairments} and any $\alpha$ satisfies
\begin{align}
\!\!\!\! \frac{\mathbb{E} \{ \|\vect{H}\|_F^2 \}  }{N_t \, \mathbb{E} \{ |h|^2 \}} & \leq& \!\!\!\! \lim_{\SNR \rightarrow 0} \mathcal{M}(\SNR) &\leq&  \frac{\mathbb{E} \{ \|\vect{H}\|_2^2 \}  }{\mathbb{E} \{ |h|^2 \}}, \qquad \quad \label{eq_refined_mg_lower} \\
M &\leq& \!\!\!\!  \lim_{\SNR \rightarrow \infty} \mathcal{M}(\SNR) &\leq& \!\!\!\! M \frac{\log_2 ( 1+\frac{N_t}{M \kappa^2} )}{\log_2 ( 1+\frac{1}{\kappa^2} )}, \label{eq_refined_mg_upper}
\end{align}
where $\|\cdot\|_F$ and $\|\cdot\|_2$ denote the Frobenius and spectral norm, respectively. The upper bounds are achieved for deterministic channels (with full rank and $\alpha=1$). The lower bounds are achieved for right-rotationally invariant channel distributions.
\end{theorem}
\begin{IEEEproof}
The low-SNR behavior is achieved by Taylor approximation: $\vect{Q}=\frac{1}{N_t}\vect{I}$ gives the lower bound, while the per-realization-optimal $\vect{Q}=\vect{u}\vect{u}^H$ (where $\vect{u}$ is the dominating eigenvector of $\vect{H}^H \vect{H}$) gives the upper bound. The high-SNR behavior follows from Theorem~\ref{theorem_upper_bound} and its corollaries.
\end{IEEEproof}

This theorem indicates that transceiver impairments have little impact on the relative MIMO gain, which is a very positive result for practical applications.
The low-SNR behavior in \eqref{eq_refined_mg_lower} is the same as for ideal transceivers (since $\SNR\vect{H} \vect{\Upsilon}_t \vect{H}^H+ \vect{I} \approx \vect{I}$), while \eqref{eq_refined_mg_upper} shows that physical MIMO channels can achieve $\mathcal{M}(\SNR)>M$ in the high-SNR regime (although ideal transceivers only can achieve $\mathcal{M}(\SNR)=M$).

\subsection{Numerical Illustrations}

\begin{figure}
\begin{center}
\includegraphics[width=1.01\columnwidth]{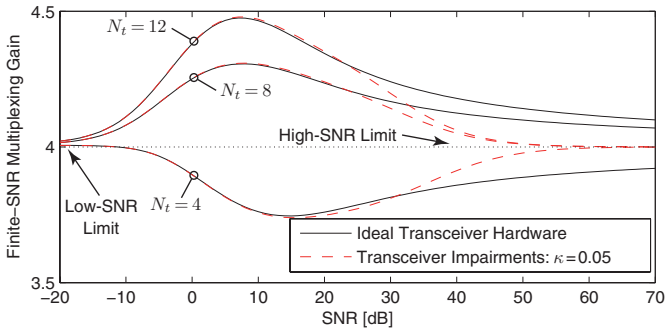}
\end{center}  \vskip-4mm
\caption{Finite-SNR multiplexing gain for an uncorrelated Rayleigh fading channel with $N_r=4$ and $N_t \geq 4$.}\label{figure_simulation_multiplexing_iid}
\end{figure}

\begin{figure}
\begin{center}
\includegraphics[width=1.01\columnwidth]{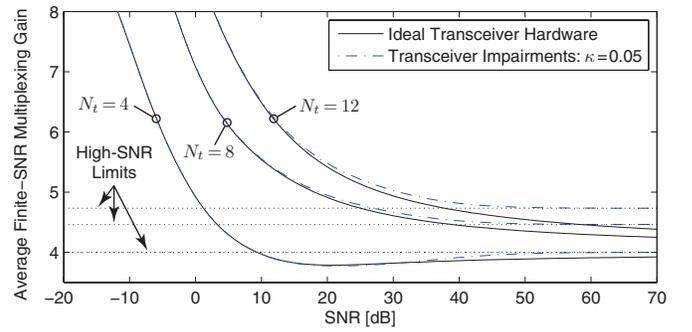}
\end{center} \vskip-4mm
\caption{Average finite-SNR multiplexing gain of deterministic channels (generated with independent $\mathcal{CN}(0,1)$-entries) with $N_r=4$ and $N_t \geq 4$.}\label{figure_simulation_multiplexing_det}
\end{figure}

The finite-SNR multiplexing gain is shown in Figs.~\ref{figure_simulation_multiplexing_iid} and \ref{figure_simulation_multiplexing_det} for uncorrelated Rayleigh fading and deterministic channels, respectively, with $N_t \in \{4, \, 8, \, 12\}$, $N_r=4$, $\kappa =0.05$, $\alpha=1$.

The limits in Theorem~\ref{theorem_finite_multiplexing_gain} are confirmed by the simulations. Although the capacity behavior is fundamentally different for physical and ideal transceivers, the finite-SNR multiplexing gain is remarkably similar---not unexpected since the asymptotic limits in Theorem~\ref{theorem_finite_multiplexing_gain} are almost the same for any level of transceiver impairments. The main difference is in the high-SNR regime, where (a) there is a faster convergence to the limits under impairments and (b) deterministic channels achieve an asymptotic gain higher than $M$ when $N_t>N_r$.

\section{Concluding Remarks}

Unlike conventional capacity analysis, the capacity of physical MIMO systems saturates in the high-SNR regime (see Theorem~\ref{theorem_upper_bound}) and the finite capacity limit is independent of the channel distribution. This fundamental result is explained by the distortion from transceiver impairments and that its power is proportional to the signal power. The classic multiplexing gain is thus zero (see Eq.~\eqref{eq_classic_multiplexing_gain}). Nevertheless, the MIMO capacity grows roughly linearly with $M=\min(N_t,N_r)$ (see Theorem~\ref{theorem_finite_multiplexing_gain}) over the whole SNR range, thus showing the encouraging result that also physical systems can achieve great gains from employing MIMO and spatial multiplexing.

Technological advances can reduce transceiver impairments, but there is currently an opposite trend towards small low-cost low-power transceivers where the inherent \emph{dirty RF effects} are inevitable and the transmission is instead adapted to them.

The point-to-point MIMO capacity limit in Theorem~\ref{theorem_upper_bound} is an upper bound for scenarios with extra constraints; for example, network MIMO, which is characterized by distributed power constraints and limited coordination both between transmit antennas and between receive antennas. The capacity in such scenarios
therefore saturates in the high-SNR regime---even in small networks where the analysis in \cite{Lozano2013a} is not applicable.

Finally, note that the finite-SNR multiplexing gain decreases when adding extra constraints \cite{Bjornson2012b} and that impairments limit the asymptotic accuracy of channel acquisition schemes.

\section*{Appendix: Proof of Theorem \ref{theorem_upper_bound}}

As a preliminary, consider any full-rank channel realization $\vect{H}$. Let $\vect{H}^H \vect{H} = \vect{U}_M \boldsymbol{\Lambda}_M \vect{U}_M^H$ denote a \emph{compact} eigendecomposition (with $\vect{U}_M \in \mathbb{C}^{N_t \times M}$, $\boldsymbol{\Lambda}_M \in \mathbb{C}^{M \times M}$; see Corollary \ref{corollary_capacity_deterministic}). The mutual information increases with $\SNR$ (since it reduces the noise term and $\log_2 \det(\cdot)$ is concave) and satisfies
\begin{align} \notag
\log_2 & \det \! \left( \vect{I} \!+\! \SNR \vect{H} \vect{Q} \vect{H}^H (  \SNR \vect{H} \vect{\Upsilon}_t \vect{H}^H \!+\! \vect{I})^{-1} \right)\\ \notag
&= \log_2 \det \! \left( \vect{I} \!+\! \SNR \vect{U}_M^H (\vect{Q}+ \vect{\Upsilon}_t) \vect{U}_M \boldsymbol{\Lambda}_M \right) \\ \notag
&\qquad \qquad - \log_2 \det \! \left( \vect{I} \!+\! \SNR \vect{U}_M^H \vect{\Upsilon}_t \vect{U}_M \boldsymbol{\Lambda}_M  \right)  \quad \rightarrow \\ \notag
  \log_2 &\det \! \left( \vect{U}_M^H(\vect{Q} \!+\! \vect{\Upsilon}_t) \vect{U}_M \boldsymbol{\Lambda}_M \right) -
\log_2 \det \! \left( \vect{U}_M^H \vect{\Upsilon}_t \vect{U}_M \boldsymbol{\Lambda}_M \right) \\ \label{eq_proof_derivation1}
& = \log_2 \det \! \left( \vect{I} + \vect{U}_M^H \vect{Q} \vect{U}_M (\vect{U}_M^H \vect{\Upsilon}_t \vect{U}_M)^{-1} \right) \\ \notag  &
 =  \log_2 \det \! \Big( \vect{I} + \vect{\Upsilon}_t^{-1/2}\vect{Q} \vect{\Upsilon}_t^{-H/2} \projUpsilon \Big)
\\[-1mm] & = \sum_{i=1}^{M} \log_2 \! \Big( 1 \!+\! \mu_i( \vect{\Upsilon}_t^{-1/2} \vect{Q} \vect{\Upsilon}_t^{-H/2} \projUpsilon)  \Big) \label{eq_proof_derivation2}
\end{align}
\noindent as $\SNR \rightarrow \infty$. The first equality follows from expanding the logarithm and from the rule $\det(\vect{I}+\vect{A} \vect{B}) = \det(\vect{I}+\vect{B}\vect{A})$. This enables letting $\SNR \rightarrow \infty$ and achieve an expression where the impact of $\boldsymbol{\Lambda}_M$ cancels out. We then identify the projection matrix $\projUpsilon = \vect{\Upsilon}_t^{H/2} \vect{U}_M (\vect{U}_M^H \vect{\Upsilon}_t \vect{U}_M)^{-1} \vect{U}_M^H \vect{\Upsilon}_t^{1/2}$ onto $\vect{U}_M^H \vect{\Upsilon}_t^{1/2}$. The $i$th strongest eigenvalue is denoted $\mu_i(\cdot)$.

As the convergence $\SNR \rightarrow \infty$ is uniform, we can achieve bounds by showing that all realizations have the same asymptotic bound.
A lower bound is given by any feasible $\vect{Q}$; we select $\vect{Q} = \frac{1}{N_t} \vect{I}$ as it gives $\vect{\Upsilon}_t = \frac{\kappa^2}{N_t} \vect{I}$
and makes \eqref{eq_proof_derivation1} independent of $\vect{H}$. Since \eqref{eq_proof_derivation2} is a Schur-concave function in the eigenvalues, an upper bound is achieved by replacing $\mu_i(\cdot)$ with the average eigenvalue $\frac{1}{M} \tr(\vect{\Upsilon}_t^{-1/2}\vect{Q} \vect{\Upsilon}_t^{-H/2} \projUpsilon) \leq \frac{1}{M} \tr(\vect{\Upsilon}_t^{-1/2}\vect{Q} \vect{\Upsilon}_t^{-H/2}) = \frac{N_t \kappa^2}{M}$, where the inequality follows from removing the projection matrix (since $\projUpsilon \preceq \vect{I}$). Note that the upper and lower bounds coincide when $N_t \leq N_r$, thus $\vect{Q} = \frac{1}{M} \vect{I}$ is asymptotically optimal in this case.

\bibliographystyle{IEEEtran}
\bibliography{IEEEabrv,letter_refs}

\end{document}